\documentclass[fleqn,12pt,twoside]{article}
\usepackage{espcrc1}

\usepackage{graphicx}
\usepackage{epsfig}
\usepackage[figuresright]{rotating}

\newcommand{\AmS}{{\protect\the\textfont2
  A\kern-.1667em\lower.5ex\hbox{M}\kern-.125emS}}

\title{Resonances in three-body systems with short and long-range interactions}

\author{E. Garrido\address[MCSD]{ Instituto de Estructura de la Materia, CSIC,
Serrano 123, E-28006 Madrid, Spain}, %
        D.V. Fedorov\address[IFA]{Department of Physics and Astronomy,
        University of Aarhus, DK-8000 Aarhus C, Denmark}, 
        and
        A.S. Jensen\addressmark[IFA]}
       
\begin{document}

\maketitle

\begin{abstract}
The complex scaling method permits calculations of few-body resonances
with the correct asymptotic behaviour using a simple box boundary
condition at a sufficiently large distance. This is also valid for
systems involving more than one charged particle.  We first apply the
method on two-body systems. Three-body systems are then investigated
by use of the (complex scaled) hyperspheric adiabatic expansion
method. The case of the 2$^+$ resonance in $^6$Be and $^6$Li is
considered. Radial wave functions are obtained showing the correct
asymptotic behaviour at intermediate values of the hyperradii, where
wave functions can be computed fully numerically.
\end{abstract}

\section{INTRODUCTION}

Use of the complex energy method permits to define resonances as poles
of the $S$-matrix in the lower half of the momentum plane. The
asymptotic behavior of the wave function is then known analytically as
given by the Hankel function of first kind when only short-range
interactions are involved, or by some combination of the regular and
irregular Coulomb functions when the Coulomb potential enters.
Resonance wave functions are then usually computed by imposing the
correct analytic asymptotic behaviour.

However the problem arises when dealing with more than two particles
and when the long-range Coulomb interaction is involved. In this case
the effective charge is not well defined, and therefore the correct
asymptotics to be imposed to the solution is not known.

The purpose of this work is to show that even in this case accurate
resonance wave functions with the correct asymptotics can be
computed. This is done by use of the complex scaling method, which
transforms the exponentially divergent resonance wave function into a
function that falls off exponentially, exactly as bound states
do. This fact permits computations of resonances using the same
numerical techniques as  for bound states. In particular, a simple
box boundary condition should be enough to obtain resonances after
complex scaling. Nothing is then needed to be known and imposed in
advance about the asymptotics, but still the correct asymptotic
behaviour is recovered.

In the following section we briefly describe the basis of the
method. In section 3 we test it in the two-body case, for which the
correct asymptotics is known also for two charged particles, and it
can then be directly compared to the numerical solution. In section 4
we apply the method for a three-body system, in particular for the
2$^+$ resonances in $^6$Be ($\alpha$+$p$+$p$) and $^6$Li
($\alpha$+$p$+$n$). The convergence and asymptotic behaviour of the
solutions are discussed. We finish in section 5 with a short summary
and the conclusions.

\section{THEORETICAL FORMULATION}

A system made of $n$ constituents ($n>1$) is usually described by a radial coordinate (relative distance
for $n$=2, hyperradius for $n$=3, $\cdots$) and $3n-4$ angles. Use of the complex energy method permits 
to obtain the resonances of such a system as poles of the $S$-matrix in the lower half
of the momentum plane. This implies that the radial wave functions of the resonances behave,
for short-range potentials, as:
\begin{equation}
f_{\mbox{\scriptsize res}}(r)\rightarrow \sqrt{r}H_\xi^{(1)}(\kappa r)
\stackrel{r\rightarrow \infty }{\longrightarrow }
e^{|\kappa| r \sin \theta_R} e^{i(|\kappa|r\cos \theta_R-\xi \pi/2-\pi/4)}
\label{eq1}
\end{equation}
where $H_\xi^{(1)}$ is the Hankel function of first kind and order $\xi$,
 and $\kappa=\sqrt{2mE/\hbar}$,
where $m$ is some normalization mass and $E=|E|e^{-i2\theta_R}$ is the $n$-body resonance energy.
When the Coulomb interaction is involved, the Hankel function is replaced by the regular and irregular
Coulomb functions, such that:
\begin{equation}
f_{\mbox{\scriptsize res}}(r)\rightarrow F_\xi(\eta,\kappa r)-i G_\xi(\eta,\kappa r)
\stackrel{r\rightarrow \infty }{\longrightarrow }
e^{|\kappa| r \sin \theta_R} e^{i(|\kappa|r\cos \theta_R-\eta \ln(2\kappa r))}
\label{eq2}
\end{equation}

The first problem to face when computing resonance wave functions is the one coming from the 
exponential divergence observed in Eqs.(\ref{eq1}) and (\ref{eq2}). For complex energy, ordinary 
continuum wave functions also diverge, and the distinction between them and the resonance wave functions 
is not trivial. 

This problem is solved by the complex scaling method, where the radial coordinates 
are rotated into the complex plane by an arbitrary angle $\theta$ ($r \rightarrow r e^{i\theta})$. 
After this transformation, the radial resonance wave functions, except for some oscillatory term, behave
asymptotically like $f_{\mbox{\scriptsize res}}(r)\stackrel{r\rightarrow \infty }{\longrightarrow }
e^{-|\kappa| r \sin(\theta-\theta_R)}$, both for short and long-range potentials.
Thus, as soon as $\theta>\theta_R$ the resonance wave functions behave as bound states, and 
they can be computed following the same numerical procedures. In particular, since 
the radial wave functions go exponentially to zero, it is possible to obtain them with
a simple box boundary condition $f(r_{max})=0$, with the only requirement that $r_{max}$
is large enough such that the wave function at that distance is orders of magnitude smaller than its maximum
value. A box boundary condition is known to work for bound states, and should also work for
resonances after a complex scaling transformation.

An additional problem arises because for more than two particles the
Coulomb charge $\eta$ (and to some extent also the index $\xi$) must
be computed numerically.

\section{TWO-BODY CASE}

A two-body system is a good test of the method, since for this particular case the correct asymptotics
is known also when the Coulomb interaction is present.

Assuming a central two-body potential $V(r)$, after complex scaling, the differential equation to be solved is
simply:
\begin{equation}
\frac{d^2}{dr^2}+\frac{2\mu}{\hbar^2}\left(Ee^{i2\theta} -e^{2i\theta}V(re^{i\theta})
 -\frac{\hbar^2}{2\mu} \frac{\ell (\ell+1)}{r^2} \right) f^{(\theta)}(r)=0
\end{equation}
where $\mu$ is the reduced mass, and $\ell$ is the relative orbital angular momentum.

As an example we take a two-body system made by an $^{15}$O core and a proton. We use a nuclear interaction
as given by the gaussian $d$-wave potential in table 1 of \cite{gar04} plus, of course, the Coulomb
repulsion.  
We then solve the previous equation by imposing $f(r_{max})$=0, with
$r_{max}$=40 fm. A solution is found at the complex energy $E=0.5-i 0.004$ MeV,
that matches well with the experimental $2^-$ resonance in $^{16}$F \cite{ajz86}.

\begin{figure}
\begin{center}
\vspace*{-1cm}
\epsfig{file=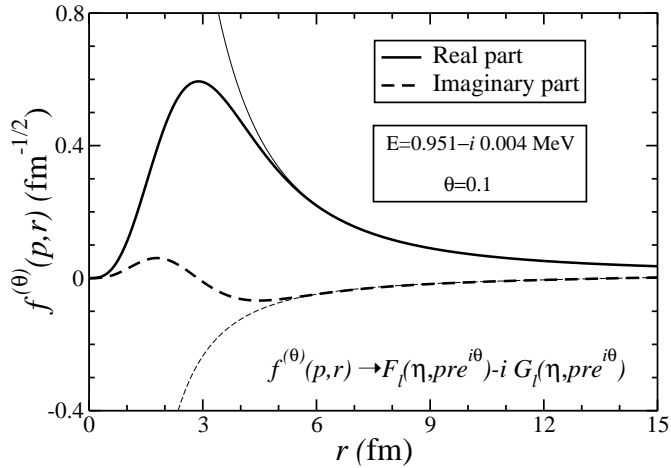,scale=0.35,angle=-90}
\end{center}
\vspace*{-1.3cm}
\caption{Thick curves: Real and imaginary parts of the complex rotated radial wave function of the
$2^-$ resonance in $^{16}$F obtained with a box boundary condition. Thin curves: Correct asymptotics of
the radial wave function as given by Eq.(\ref{eq2}).}
\label{fig1}
\end{figure}

In Fig.\ref{fig1} the thick-solid and thick-dashed curves show the
real and imaginary parts of the complex rotated radial wave function
of the resonance.  As mentioned above, in this case the asymptotics is
well known, as is given by Eq.(\ref{eq2}) with $\xi=\ell$ and
$\eta=\frac{\mu c}{p} Z_1 Z_2 \alpha$, where $Z_1$ and $Z_2$ are the
charges, $\alpha$ is the fine structure constant, and $p=\sqrt{2\mu
E/\hbar}$. This asymptotic behaviour is shown in Fig.\ref{fig1} by the
corresponding thin curves. We can see that the numerical radial wave
functions, obtained with a simple box boundary condition, reproduce
the correct asymptotic behaviour.

\section{THREE-BODY CASE}

After testing the method for two-body systems, we now use the same procedure for three-body systems,
for which the wave functions are computed by using the hyperspheric adiabatic expansion method \cite{nie01}.
In this method the different radial wave functions are obtained by solving the coupled set of differential
equations:
\begin{equation}
\left(
-\frac{d^2}{d\rho^2}-\frac{2mE}{\hbar^2}+\frac{\lambda_n(\rho)+\frac{15}{4}}{\rho^2}
-Q_{nn}\right) f_n(\rho)+ 
\sum_{n^\prime\neq n} \left(-2P_{nn^\prime}(\rho)\frac{d}{d\rho}-Q_{nn^\prime}(\rho)
\right)f_{n^\prime}(\rho)=0
\label{eq4}
\end{equation}
where $\rho$ is the hyperradius, $m$ is a normalization mass, and
$P_{nn^\prime}$ and $Q_{nn^\prime}$ are the non-adiabatic terms
coupling the different radial wave functions \cite{nie01}.

In the previous equations the key quantities are the effective potentials $\lambda_n(\rho)$ that
are obtained as the eigenvalues of the angular part of the Faddeev equations \cite{nie01}. Accurate
calculation of these eigenvalues is the first and essential requirement needed to obtain reliable
radial wave functions.

\subsection{2$^+$ resonances in $^6$Be and $^6$Li}

The main properties of $^6$He are accurately described when treating
this nucleus as a three-body system made by an $\alpha$-particle and
two neutrons. Since both the alpha-neutron and the neutron-neutron
interactions are well known, this nucleus appears as an almost perfect
test for all the available numerical three-body methods.

For the same reason, $^6$Be ($\alpha$+$p$+$p$) and $^6$Li
($\alpha$+$n$+$p$) are specially appropriate to investigate three-body
systems when more than one particle is charged. In particular, it is
well established that $^6$He has a 2$^+$ resonance with energy and
width ($E_R$,$\Gamma_R$)=(0.83,0.11) MeV. The same analog $2^+$ state
has been found in $^6$Li and $^6$Be at (1.67,0.54) MeV and (3.04,1.16)
MeV, respectively \cite{ajz88}. Realistic detailed calculations using
the (complex scaled) adiabatic expansion method concerning the 2$^+$
resonance in $^6$He can be found in
\cite{fed03,gar06}.

\begin{figure}[htb]
\begin{minipage}[t]{80mm}
\vspace*{-1cm}
\epsfig{file=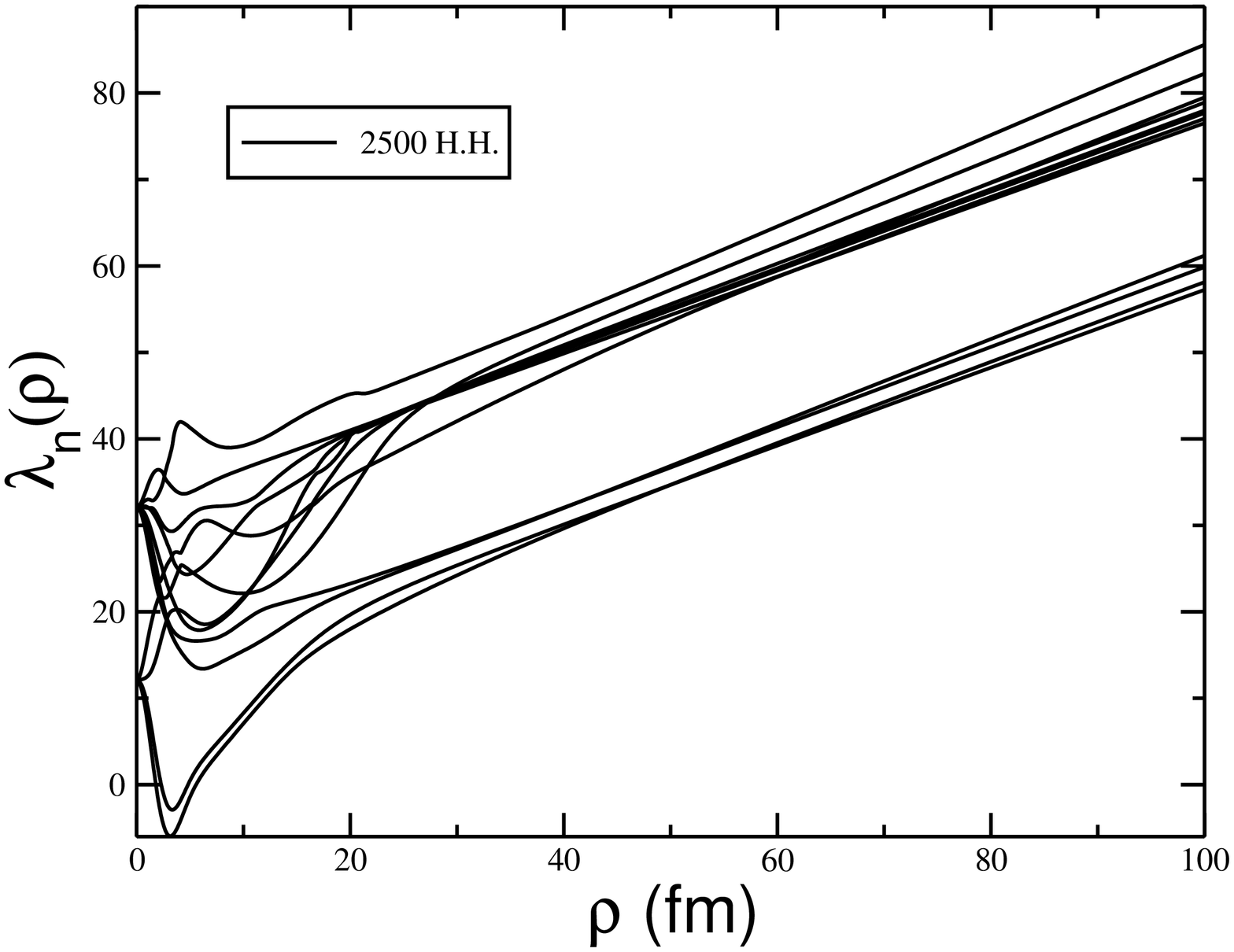,scale=0.35,angle=-90}
\vspace*{-1.3cm}
\caption{Converged $\lambda$-functions (see Eq.(\ref{eq4})) included in the calculation of the
2$^+$ resonance in $^6$Be.}
\label{fig2}
\end{minipage}
\hspace{\fill}
\begin{minipage}[t]{75mm}
\vspace*{-1cm}
\epsfig{file=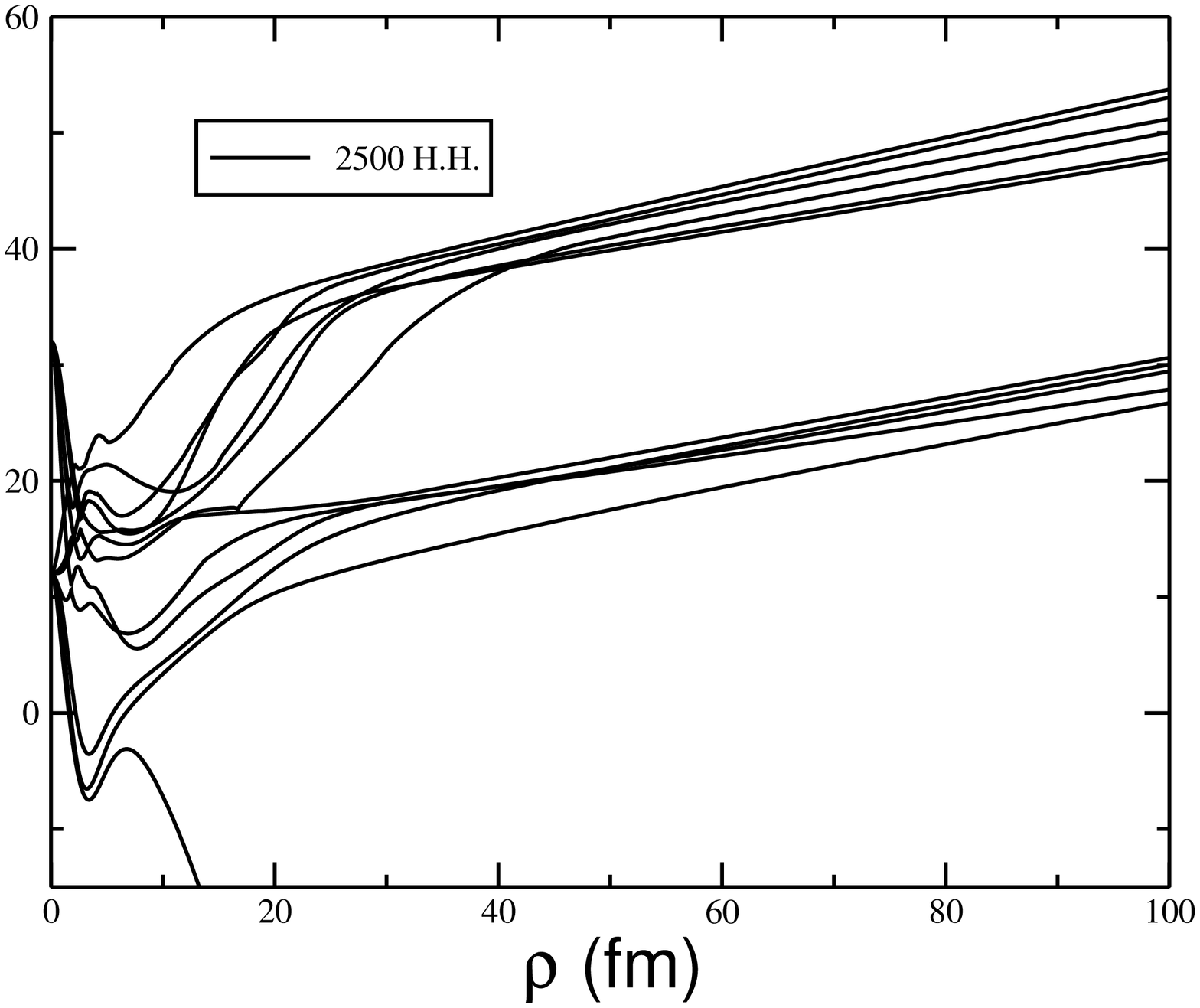,scale=0.35,angle=-90}
\vspace*{-1.3cm}
\caption{The same as in Fig.\ref{fig2} for the 2$^+$ resonance in $^6$Li. }
\label{fig3}
\end{minipage}
\end{figure}

In Figs. \ref{fig2} and \ref{fig3} we show the $\lambda$-functions
used in the calculation of the 2$^+$ resonance in $^6$Be and $^6$Li,
respectively. Partial waves with $\ell_x$ and $\ell_y$ up to 10 have
been included. A maximum value of the hypermomentum
($K=2n+\ell_x+\ell_y$) equal to 20 has been used for all the
components, but for the most relevant ones the maximum value of $K$
has been increased up to 200 for $s$-waves, 90 for $p$-waves, and 60
for $d$-waves. In total, slightly more than 2500 hyperspherical
harmonics have been included in the calculation. The computed
$\lambda$'s have then converged at least up to 100 fm. A reduction by
a factor of 2 in the basis size produces $\lambda$-functions
indistinguishable from the ones shown in the figures.

It is important to note that $^6$Li, contrary to $^6$Be and $^6$He, is
not borromean (the neutron and the proton can bind into
deuteron). This is actually revealed by the lowest $\lambda$ in
Fig.\ref{fig3}, that diverges parabolically to $-\infty$. Therefore,
the calculation for $^6$Li requires additional components that are
forbidden in $^6$Be and $^6$He. These components correspond to the
ones with zero isospin in the neutron-proton channel.

Once the $\lambda$-functions have been computed, the remaining step is
to solve the coupled set of differential equations
(\ref{eq4}). According to the discussion in the previous section,
after complex scaling a box boundary condition $f_n(\rho_{max})$=0
should be enough to obtain accurate three-body resonance wave
functions. However, as shown in Figs.\ref{fig2} and \ref{fig3} the
effective potentials entering in Eq.(\ref{eq4}) have been accurately
computed only up to $\rho_{max}$=100 fm, that is too little to expect
the box boundary condition to work. Furthermore, accurate computation
of the $\lambda$'s up to a $\rho_{max}$ value several times larger is
too expensive from the computation time point of view.

\begin{figure}[htb]
\begin{minipage}[t]{80mm}
\vspace*{-1cm}
\epsfig{file=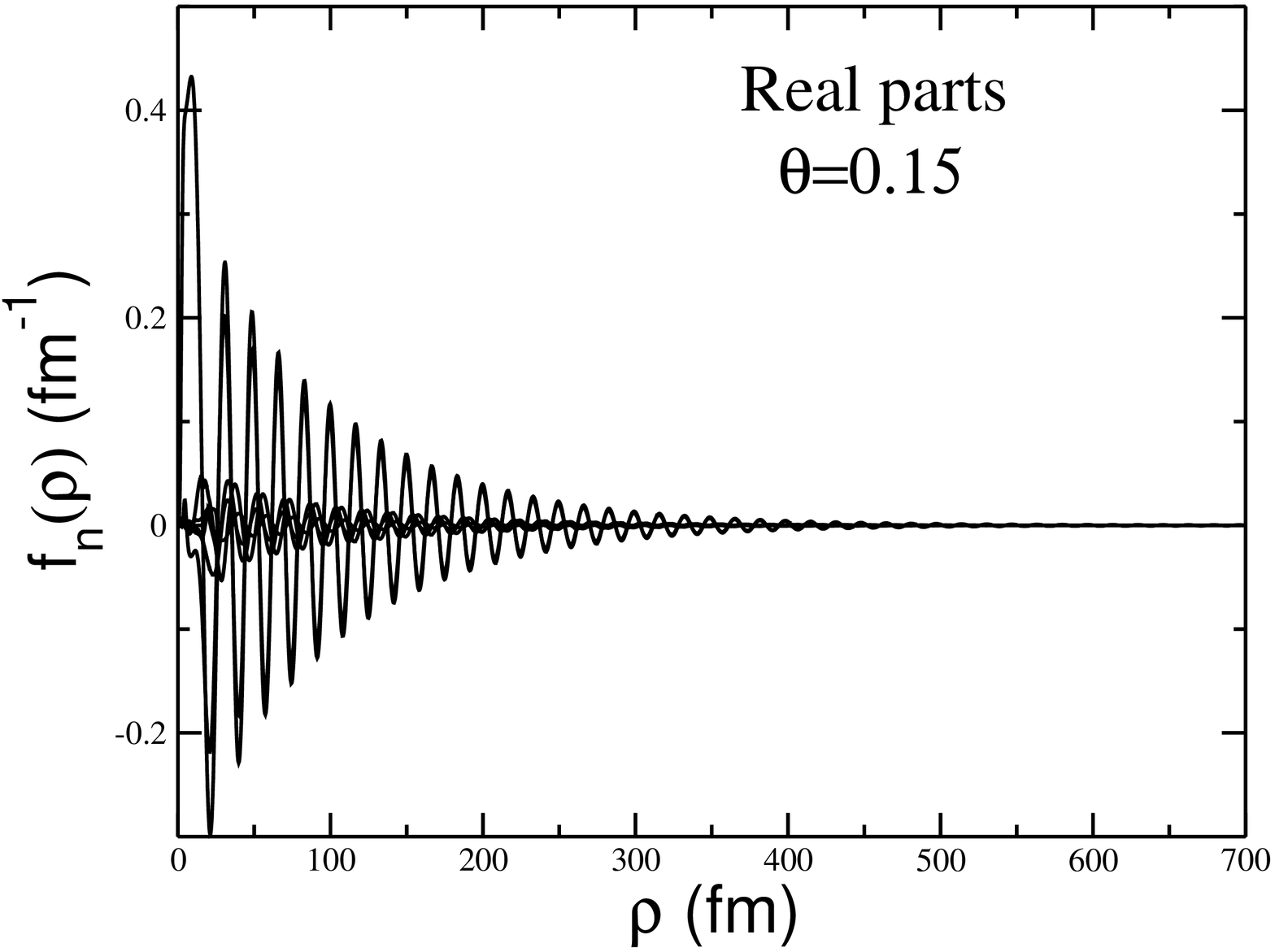,scale=0.35,angle=-90}
\vspace*{-1.3cm}
\caption{Real parts of the radial resonance wave functions of the 2$^+$ resonance in $^6$Be.}
\label{fig4}
\end{minipage}
\hspace{\fill}
\begin{minipage}[t]{75mm}
\vspace*{-1cm}
\epsfig{file=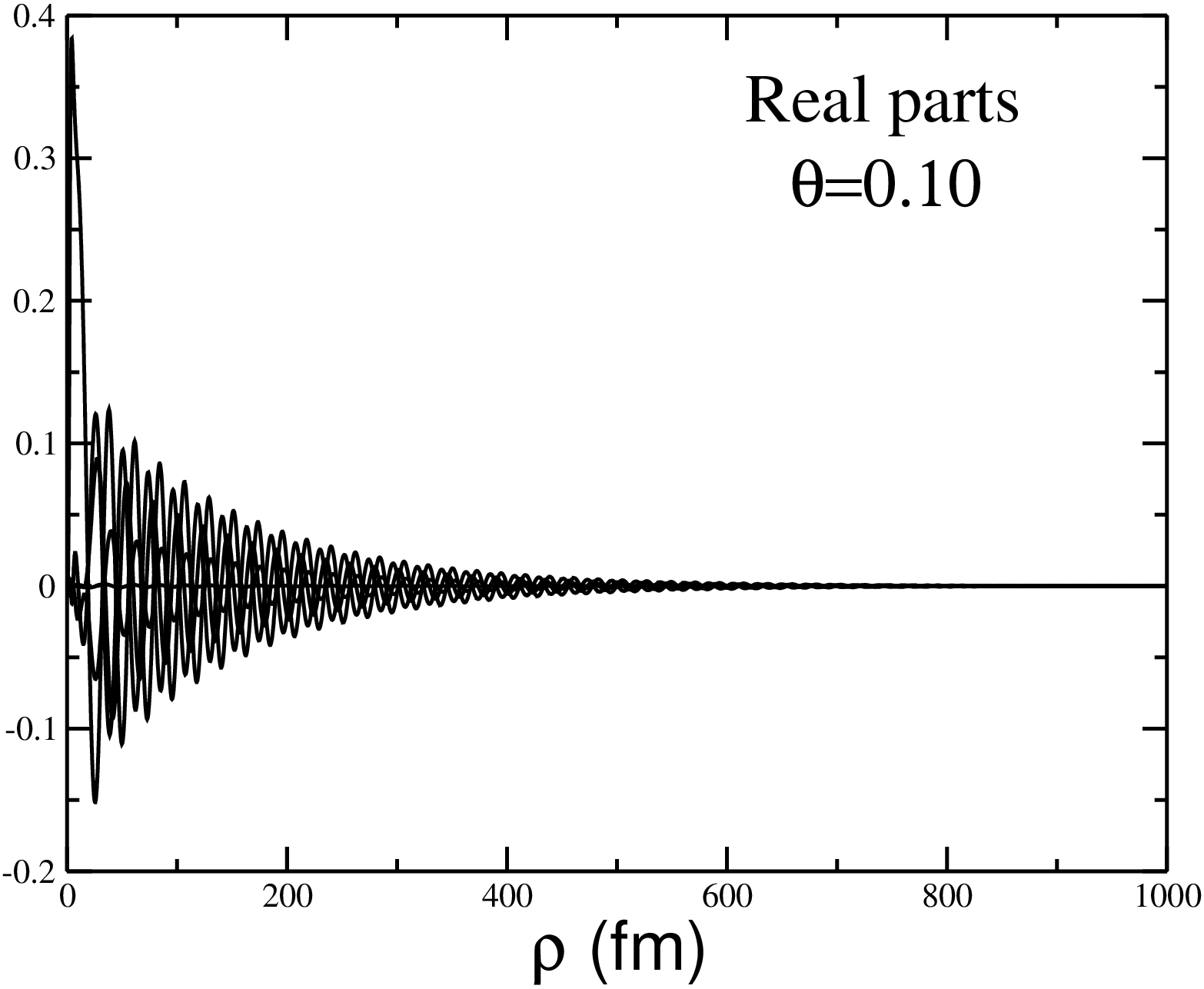,scale=0.35,angle=-90}
\vspace*{-1.3cm}
\caption{The same as in Fig.\ref{fig4} for the 2$^+$ resonance in $^6$Li. }
\label{fig5}
\end{minipage}
\end{figure}

It is not difficult to see \cite{nie01} that at large distances the  $\lambda$-functions go linearly 
with $\rho$ (this is actually obvious from Figs. \ref{fig2} and \ref{fig3}). Also the $Q_{nn}$ functions 
entering in Eq.(\ref{eq4}) go like $1/\rho^2$, while $P_{nn^\prime}$,and $Q_{nn^\prime}$ ($n\neq n^\prime$)
go to zero faster. We have then, for $\rho$ larger than 100 fm, used extrapolations of the $\lambda$'s,
$P$'s and $Q$'s in Eq.(\ref{eq4}) according to 
$\lambda_n(\rho)=A_{\lambda_n} \rho + B_{\lambda_n} + C_{\lambda_n}/\rho$,
$Q_{nn}(\rho)=A_{Q_{nn}}/\rho^2 + B_{Q_{nn}}/\rho^3 + C_{Q_{nn}}/\rho^4$, and as
$A_{P,Q}/\rho^3 + B_{P,Q}/\rho^4 + C_{P,Q}/\rho^5$ for the non-diagonal $P$'s and $Q$'s. We have then 
solved the coupled equations (\ref{eq4}) using the numerical $\lambda$'s in Figs. \ref{fig2} and \ref{fig3}
up to 100 fm, and the extrapolations for larger $\rho$'s. In Figs. \ref{fig4} and \ref{fig5} we show
the real parts of 
the complex scaled radial wave functions obtained in this way for the $2^+$ resonance in $^6$Be and $^6$Li, 
respectively. For $^6$Be we use a complex scaling angle of 0.15 rads, and $\rho_{max}$=700 fm is enough
to obtain a 2$^+$ resonance with energy and width ($E_R$,$\Gamma_R$)=(2.94,1.45) MeV, that agree rather
well with the experimental values \cite{ajz88}. For $^6$Li the complex scaling angle is 0.10 rads and
$\rho_{max}$=1000 fm. We obtain ($E_R$,$\Gamma_R$)=(1.67,0.50) MeV, that also agrees well
with the experiment \cite{ajz88}. From the figures it is now clear that 100 fm is certainly not large 
enough to impose a box boundary condition at that distance.

\begin{figure}[htb]
\begin{minipage}[t]{80mm}
\vspace*{-1cm}
\epsfig{file=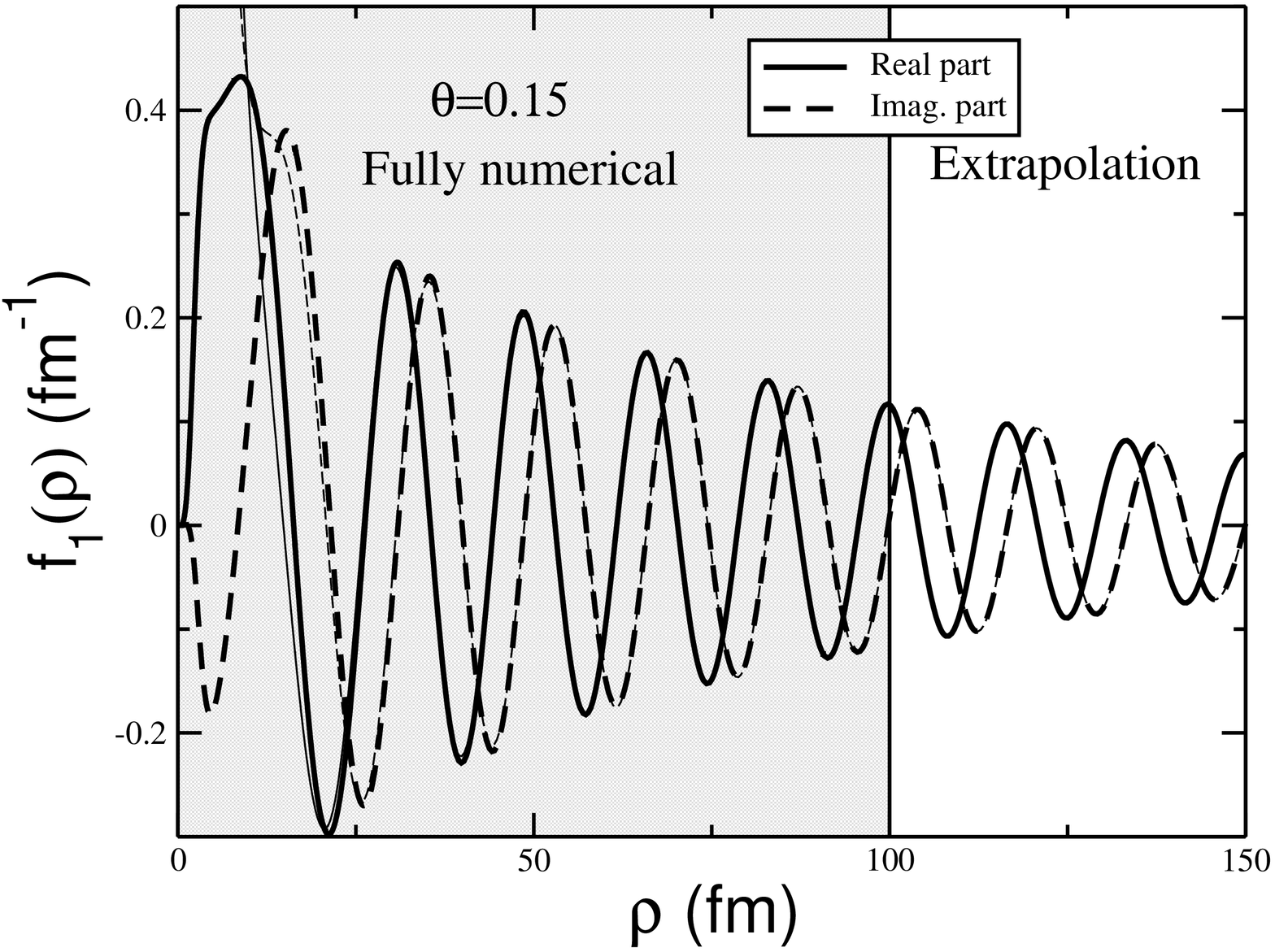,scale=0.35,angle=-90}
\vspace*{-1.3cm}
\caption{First computed radial wave function (thick curves) for the 2$^+$ resonance in $^6$Be and the
corresponding asymptotics (thin curves).}
\label{fig6}
\end{minipage}
\hspace{\fill}
\begin{minipage}[t]{75mm}
\vspace*{-1cm}
\epsfig{file=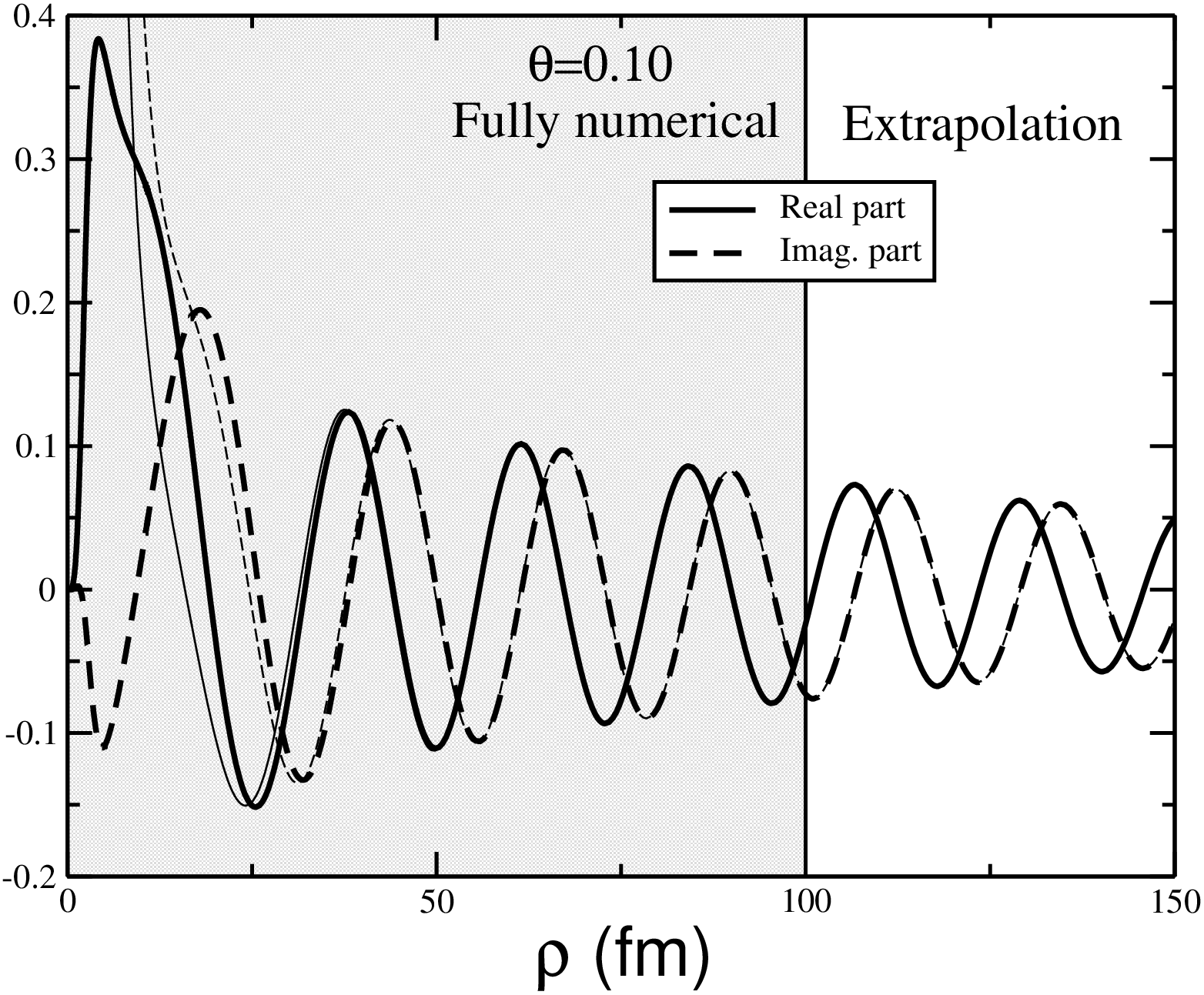,scale=0.35,angle=-90}
\vspace*{-1.3cm}
\caption{The same as in Fig.\ref{fig6} for the 2$^+$ resonance in $^6$Li. }
\label{fig7}
\end{minipage}
\end{figure}
 
In the extrapolated region ($\rho>100$ fm) the effective potentials entering in Eq.(\ref{eq4}) go like
$A/\rho+B/\rho^2+\cdots$, where the $A$ and $B$ coefficients are computed numerically. 
For such a kind of potential the solutions are known to go asymptotically as in Eq.(\ref{eq2}), 
where $\xi$ and $\eta$ are easily related to 
$A$ and $B$. In Figs. \ref{fig6} and \ref{fig7} we show the first radial wave function for 
the $2^+$ resonance in $^6$Be and $^6$Li, respectively. The thick curves are the numerical solutions, while the
thin curves show the analytic asymptotic behaviour as given by Eq.(\ref{eq2}). As seen in both figures the
matching with the numerical calculations is very good in the ``extrapolated" region, as expected. The 
remarkable fact is that this agreement is also excellent already around 50 fm, in the region where the purely 
numerical effective potentials are used. 

\begin{figure}[htb]
\begin{minipage}[t]{80mm}
\vspace*{-1cm}
\epsfig{file=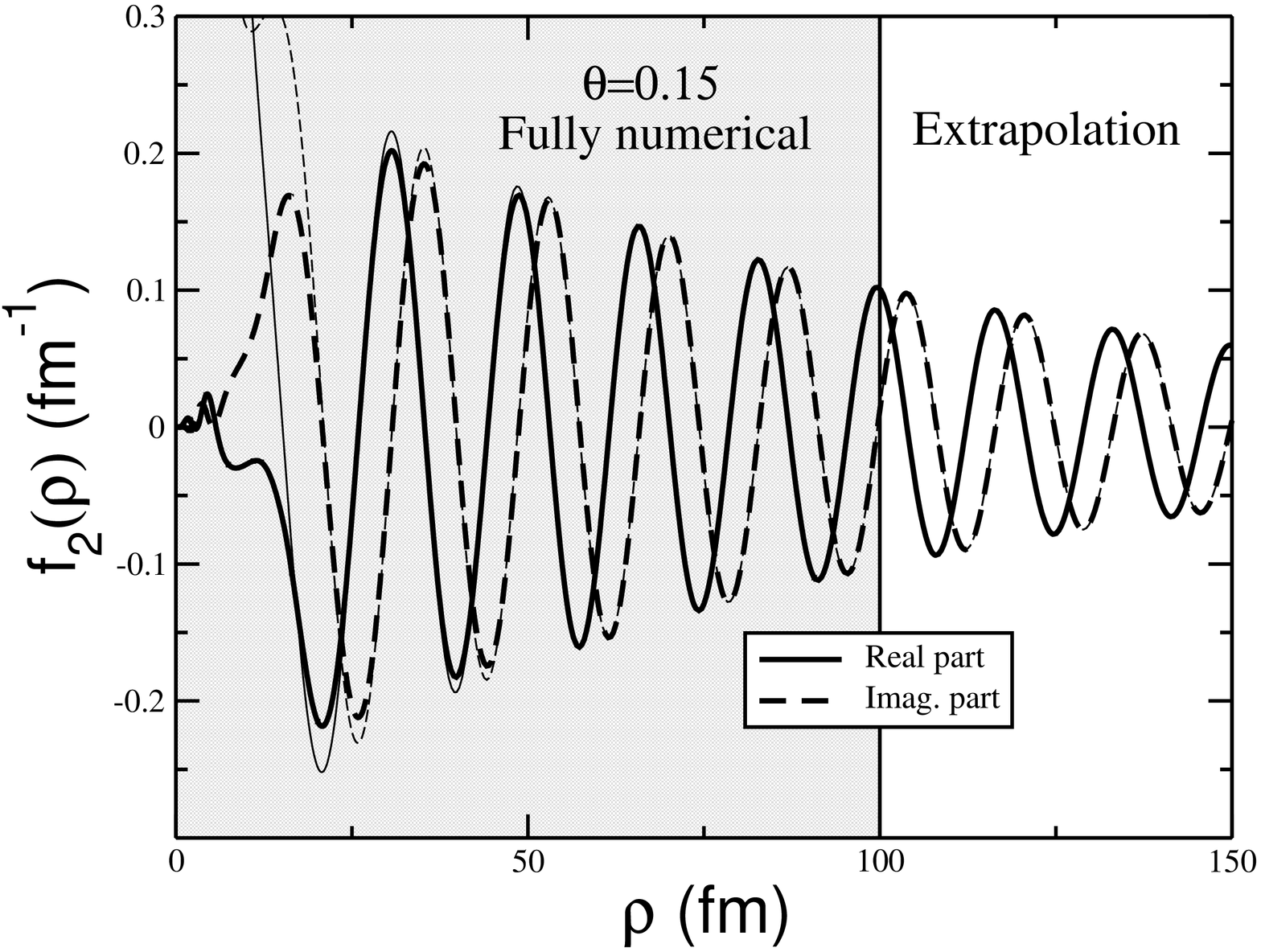,scale=0.35,angle=-90}
\vspace*{-1.3cm}
\caption{Second computed radial wave function (thick curves) for the 2$^+$ resonance in $^6$Be and the
corresponding asymptotics (thin curves).}
\label{fig8}
\end{minipage}
\hspace{\fill}
\begin{minipage}[t]{75mm}
\vspace*{-1cm}
\epsfig{file=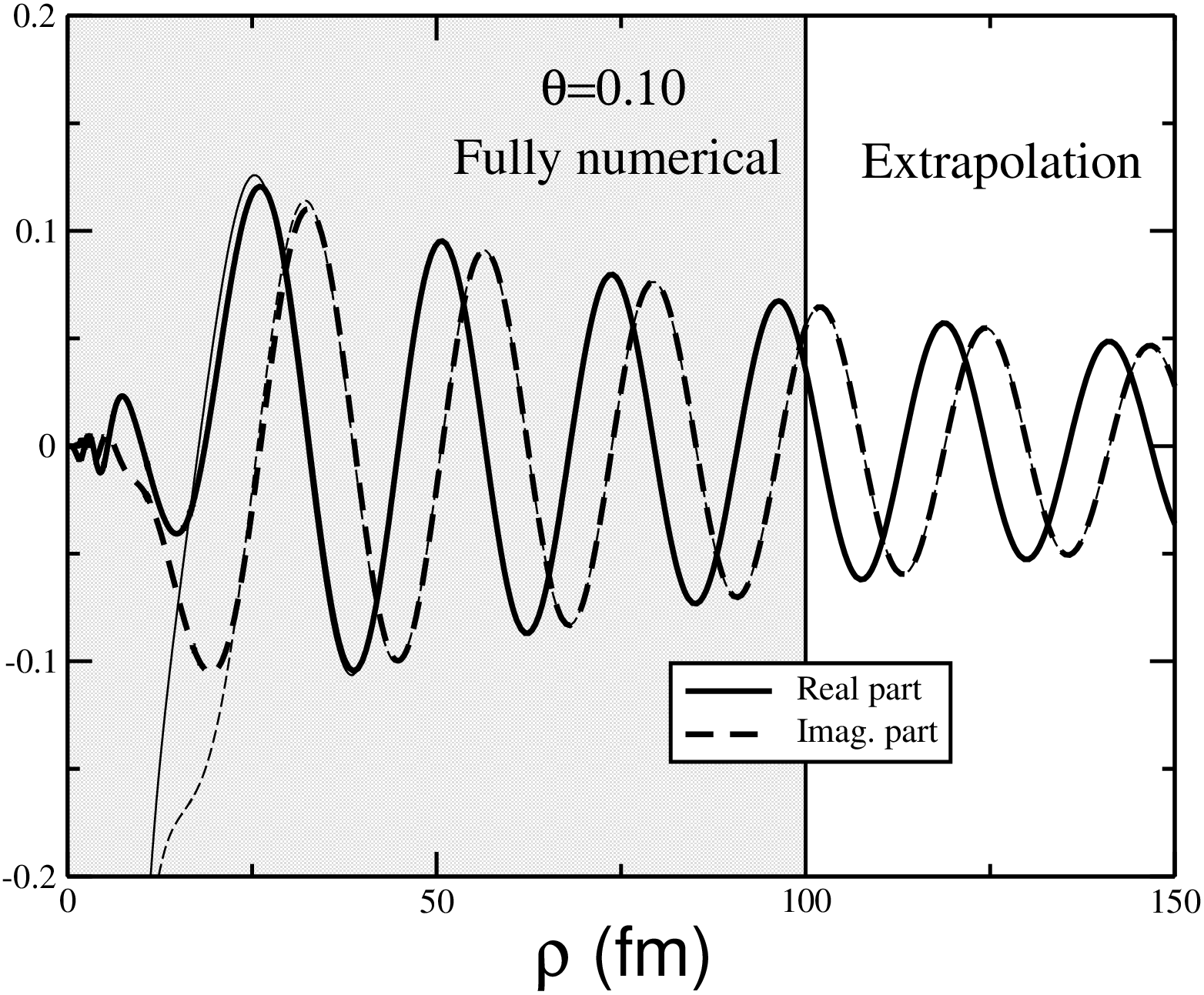,scale=0.35,angle=-90}
\vspace*{-1.3cm}
\caption{The same as in Fig.\ref{fig8} for the 2$^+$ resonance in $^6$Li. }
\label{fig9}
\end{minipage}
\end{figure}

In Figs. \ref{fig8} and \ref{fig9} we observe how the same behaviour
appears for the second radial wave function. Similar results are
also found for the other radial functions. Therefore the numerical
resonance radial wave functions obtained with the converged effective
potentials shown in Figs. \ref{fig2} and \ref{fig3}, are consistent
with the extrapolations used for the $\lambda$'s, $P$'s and $Q$'s in
Eq.(\ref{eq4}), and also these solutions have converged to the
expected asymptotic behaviour already at $\rho$-values around 50 fm,
in the region where full accurate numerical calculations can be
performed.

\section{SUMMARY AND CONCLUSIONS}

After a complex scaling transformation resonance wave functions behave
asymptotically as bound states, i.e. falls off exponentially. In this
work we exploit this fact to obtain resonances simply by using a box
boundary condition at a sufficiently large distance.  In this way
prior knowledge of the correct asymptotic behaviour of the wave
function is not required, permitting then to compute resonances for
systems for which this asymptotics is not known, in particular for
three-body systems involving more than one charged particle.

After testing the two-body case (for which the correct asymptotics is known), we have investigated
the case of the 2$^+$ resonance in $^6$Be and $^6$Li. These two nuclei are specially appropriate, since
the uncertainties coming from the two-body interactions are small. The (complex scaled) hyperspheric
adiabatic expansion method is used. Accurate and converged effective potentials are obtained up to 
$\rho_{max}$=100 fm, and beyond this distance extrapolations are used. These extrapolations permit
to know the correct asymptotic behaviour, that should be the right one at least for distances much 
larger than $\rho_{max}$.

When solving the radial equations with a box boundary condition at a large value of $\rho$ (700 fm
for $^6$Be and 1000 fm for $^6$Li) we have found that at rather modest distances ($\rho\approx$50 fm) 
the radial wave functions already match with the asymptotic behaviour obtained from the extrapolated
effective potentials. This means that the numerical effective potentials obtained for $\rho<\rho_{max}$ 
are consistent with the extrapolations used to solve the radial equations. Furthermore, use of these
numerical potentials is enough to obtain radial resonance wave functions that have already reached the 
asymptotic behaviour. This implies that observables related to the asymptotics of the
wave functions can be safely computed using the resonance wave functions obtained by this procedure. An 
example is the energy distributions of the fragments after decay of the resonance. The experimental
energy distributions are related to the energy distributions of the fragments at distances where 
the correct asymptotics has been reached \cite{fed04}.


\begin{thebibliography}{9}
\bibitem{gar04} E. Garrido, D.V. Fedorov, A.S.~Jensen,
Nucl. Phys. A 733 (2004) 85.
\bibitem{ajz86} F. Ajzenberg-Selove, Nucl. Phys. A 460 (1986) 1.
\bibitem{nie01} E. Nielsen, D.V. Fedorov, A.S. Jensen, E. Garrido,
Phys. Rep. 347 (2001) 373.
\bibitem{ajz88} F. Ajzenberg-Selove, Nucl. Phys. A 490 (1988) 1.
\bibitem{fed03} D.V. Fedorov, E. Garrido, A.S. Jensen, Few-Body Syst. 33 (2003) 153.
\bibitem{gar06} E. Garrido, D.V. Fedorov, A.S.~Jensen, H.O.U. Fynbo,
Nucl. Phys. A 766 (2006) 74.
\bibitem{fed04} D.V. Fedorov, H.O.U. Fynbo, E. Garrido, A.S.~Jensen, 
Few-Body Syst. 34 (2004) 33.
\end{thebibliography}
\end{document}